\documentclass[usletter,12pt,leqno]{article}

\usepackage{graphicx}
\usepackage{amsmath,amsthm,amssymb,mathrsfs,mathtools}
\usepackage[OT1]{fontenc}
\usepackage{etoc}
\usepackage{float}
\usepackage{fancyhdr}
\def\subsectiontitle{}
\def\subsubsectiontitle{}
\usepackage{tikz}
\usetikzlibrary{automata, positioning}

\usepackage[noamsmath]{kpfonts}
\usepackage{zi4}

\usepackage{enumitem}
\usepackage[margin=1in]{geometry}
\usepackage{breakcites}
\usepackage{hyperref}
\usepackage[nameinlink,noabbrev,sort,capitalise]{cleveref}
\makeatletter
\def\ps@pprintTitle{%
	\let\@oddhead\@empty
	\let\@evenhead\@empty
	\def\@oddfoot{\emph{Very preliminary version}\hfill\emph{This draft: \today}}%
	\let\@evenfoot\@oddfoot}
\makeatother
\usepackage{etoolbox}
\patchcmd{\pprintMaketitle}
{\fi\hrule}% the second rule
{\fi\ifvoid\extrainfobox\else\unvbox\extrainfobox\par\vski\succ_10pt\fi\hrule}
{}{}

\newsavebox\extrainfobox

\usepackage{comment}
\usepackage{multirow}
\usepackage{multicol}
\usepackage{subcaption}

\crefname{problem}{Problem}{Problems}

\allowdisplaybreaks

\let\oldfootnote\footnote
\renewcommand\footnote[1]{\oldfootnote{\hspace{.4mm}#1}}
\makeatletter
\renewenvironment{proof}[1][\proofname] {\par\pushQED{\qed}\normalfont\topsep6\p@\@plus6\p@\relax\trivlist\item[\hskip\labelsep\bfseries#1\@addpunct{.}]\ignorespaces}{\popQED\endtrivlist\@endpefalse}
\makeatother
\let\oldFootnote\footnote
\newcommand\nextToken\relax

\renewcommand\footnote[1]{%
	\oldFootnote{#1}\futurelet\nextToken\isFootnote}

\newcommand\isFootnote{%
	\ifx\footnote\nextToken\textsuperscript{,}\fi}
\usepackage{blkarray}
\usepackage{graphicx,pgfpages,epsfig,multirow,lscape}
\usepackage{epic}
\usepackage{setspace}
\usepackage{tikz}
\usetikzlibrary{arrows,decorations,decorations.pathreplacing,calc,matrix}
\usepackage{natbib}
\DeclareFontFamily{U}{mathb}{\hyphenchar\font45}
\DeclareFontShape{U}{mathb}{m}{n}{
	<-6> mathb5 <6-7> mathb6 <7-8> mathb7
	<8-9> mathb8 <9-10> mathb9
	<10-12> mathb10 <12-> mathb12
}{}
\DeclareSymbolFont{mathb}{U}{mathb}{m}{n}
\DeclareMathSymbol{\llcurly}{\mathrel}{mathb}{"CE}
\DeclareMathSymbol{\ggcurly}{\mathrel}{mathb}{"CF}

\def \D{\mathcal{D}}

\usepackage{xcolor}

\def\th@colored{%
	\normalfont
	\thm@headfont{\color{blue}\bfseries}%
	\thm@notefont{\bfseries}%
	\thm@bodyfont{\color{black}}%
}
\makeatother
\theoremstyle{colored}

\newtheorem{definition}{Definition}

\newtheorem{theorem}{Theorem}
\newtheorem*{theorem*}{Theorem}
\newtheorem{proposition}{Proposition}
\newtheorem{lemma}{Lemma}
\newtheorem{example}{Example}

\hypersetup{colorlinks = true}
\hypersetup{allcolors=blue}

\usepackage{setspace}
\newcommand{\hide}[1]{}

\begin{document}
	
	\title{Local unanimity in Shapley-Scarf housing markets\thanks{Xinquan Hu thanks the support of the National Science Foundation of China (72394391). Jun Zhang thanks the support of the National Science Foundation of China (72122009, 72033004) and the Wu Jiapei Foundation of the China Information Economics Society (E21103567).}
    }
	
	\author{Xinquan Hu\thanks{School of Business, Xiangtan University. Email: xinquan.hu.eco@gmail.com.} \qquad Jun Zhang\thanks{Institute for Social and Economic Research, Nanjing Audit University. Email: zhangjun404@gmail.com.}
    }
	
	\date{November 6, 2025}
	
	\maketitle
	
	\begin{abstract}
		In the housing market model introduced by \citet{shapley1974cores}, we propose a new axiom called \textit{local unanimity}, which extends the unanimity condition widely used in social choice theory. It applies the unanimity condition to any subset of agents in the model who unanimously agree on the best exchange of their endowments. Building on this axiom, we provide several concise characterizations of the Top Trading Cycles (TTC) mechanism under both strict and weak preference domains.
	\end{abstract}
	
	\vspace{1cm}
	
	\noindent \textbf{Keywords}: housing market model; top trading cycles; local unanimity; strategy-proofness
	
	\vspace{.2cm}
	
	\noindent \textbf{JEL Classification}: C71, C78, D78
	
	\thispagestyle{empty}
	\setcounter{page}{0}
	
	\clearpage
	
	\section{Introduction}
	
	Many market design problems can be formulated as the allocation of indivisible objects to unit-demand agents without involving monetary transfers. A classical model in this field is the \textit{housing market} model introduced by \cite{shapley1974cores}, which considers the exchange of endowments among a finite set of agents, each endowed with a distinct object and demanding one object. The central question is to identify a desirable mechanism that determines how agents exchange their endowments based on their reported preferences. \citeauthor{shapley1974cores} propose the top trading cycles (TTC) mechanism (and attribute it to David Gale), which has become one of the most influential mechanisms in market design theory and applications. When agents' preferences are strict, TTC possesses desirable properties such as Pareto efficiency, individual rationality, and strategy-proofness. A seminal characterization of TTC is due to \cite{ma1994strategy}, who proves that TTC is the only mechanism satisfying these three properties. The purpose of this paper is to provide new characterizations of TTC grounded in an intuitive axiom.

	Our new axiom is a natural adaptation of the \textit{unanimity} condition widely used in social choice theory \citep{arrow1951social,gibbard1973manipulation,satterthwaite1975strategy} to the housing market model. As one of the oldest and most fundamental normative principles for aggregating individual preferences into a collective decision, unanimity requires that if agents unanimously agree that a specific social outcome is best, then that outcome should be selected. If viewing the housing market model as a social choice problem and applying this condition, unanimity would require that if all agents agree that there is a best allocation of their objects, then this allocation should be implemented. Under this interpretation, \cite{takamiya2001coalition} and \cite{feng2024characterizing} have respectively used unanimity along with other axioms to characterize TTC and its extension to the multiple-type housing market model.\footnote{To be precise, \cite{takamiya2001coalition} characterizes TTC by individual rationality, ontoness, and group strategy-proofness. Ontoness is weaker than unanimity, but they are equivalent in the presence of group strategy-proofness. Moreover, ontoness and group strategy-proofness together imply Pareto efficiency, so \citeauthor{takamiya2001coalition}'s result is implied by \citeauthor{ma1994strategy}'s.} % A unanimously best allocation is one in which all agents receive their most preferred objects, which can exist only when their most preferred objects are distinct. 
	However, we note that unlike the ``public good'' feature of social choice problems, the housing market model represents a ``private good'' economy: as each agent initially owns a distinct object, every subset of agents can be viewed as a valid economy that is structurally identical to the original economy. Therefore, if unanimity can be applied to the original economy, it is also natural to apply the condition to any subset of agents. This motivates us to define \textit{local unanimity}, which requires that if any subset of agents unanimously agree that an exchange of their endowments among themselves gives them the best outcome among all possible allocations in the original economy, this exchange should be implemented. % Our main results show that local unanimity and (group) strategy-proofness characterize TTC.

	Local unanimity is intuitive for the housing market model, and it is particularly appealing for understanding TTC, because it captures the central feature of the TTC procedure. In each step, if viewing the remaining agents with their endowments as a reduced economy, then whenever a group of agents can receive their most preferred objects by exchanging their endowments, TTC implements that exchange. Thus, TTC is a procedure that iteratively applies local unanimity to the reduced economy in each step. For direct mechanisms that are defined as functions from preference profiles to allocations, any mechanism satisfying local unanimity must ``coincide with the first step of TTC''. Our results show that, together with incentive compatibility or other axioms, local unanimity fully characterizes the outcome of TTC. Compared with existing characterizations of TTC, our results are more concise by using fewer axioms. The proofs are not complicated and, in fact, are simple in some results. We view these as advantages of our results.

	We first obtain two characterizations of TTC in the housing market model with a fixed population. Early characterizations of TTC are typically derived on the domain of unrestricted strict preferences, whereas recent studies have turned to restricted domains and gained new insights. To clarify the scope of our results and distinguish them from existing ones, we explicitly specify the conditions on the preference domains under which our results hold. 	
	We consider two such conditions: \textit{Top-one Richness} requires that agents be able to report any object as their first choice, while \textit{Top-two Richness} requires that agents be able to report any two objects as their top two choices and rank them in any order. If there are $ n $ objects, the unrestricted strict preference domain contains $ n! $ elements, whereas a domain satisfying Top-one Richness may contain as few as $ n $ elements, and a domain satisfying Top-two Richness may contain as few as $ n(n-1) $ elements. 
	
	\autoref{thm:top-one} proves that, on any domain satisfying Top-one Richness, TTC is the unique mechanism satisfying local unanimity and group strategy-proofness. When there are only three agents, group strategy-proofness can be replaced by strategy-proofness (\autoref{prop:three-agents}), However, when there are at least four agents, this relaxation is no longer possible. We construct a mechanism satisfying local unanimity and strategy-proofness but violating group strategy-proofness on the domain of single-peaked preferences (\autoref{example:single-peaked}). The single-peaked domain satisfies Top-one Richness.  On this domain, \cite{ma1994strategy}'s characterization of TTC no longer holds because \cite{bade2019matching} finds a distinct mechanism that satisfies his axioms. Hence, \autoref{thm:top-one} is independent of \citeauthor{ma1994strategy}'s result. \citeauthor{bade2019matching}'s mechanism is, in fact, group strategy-proof. So, by \autoref{thm:top-one}, \citeauthor{bade2019matching}'s mechanism violates local unanimity. We further show that the Top-one Richness condition is indispensable for \autoref{thm:top-one}. We construct a mechanism satisfying local unanimity and group strategy-proofness but differing from TTC on the domain of single-dipped preferences (\autoref{example:single-dipped}). The single-dipped domain violates Top-one Richness.

	\autoref{thm:top-two} proves that, on any domain satisfying Top-two Richness, TTC is the unique mechanism satisfying local unanimity and strategy-proofness. Compared with \autoref{thm:top-one}, this result weakens the incentive compatibility axiom but strengthens the condition on the preference domain. It is independent of \citeauthor{ma1994strategy}'s result: in the absence of strategy-proofness, local unanimity does not imply individual rationality and Pareto efficiency, and the converse also fails. The Top-two Richness condition is indispensable for \autoref{thm:top-two} because, as mentioned above, we can construct a mechanism satisfying local unanimity and strategy-proofness but differing from TTC on the single-peaked domain.
	
	Under a fixed population, local unanimity has limited force, as it imposes no restrictions on agents who do not have a unanimously best exchange of their objects. This is where (group) strategy-proofness plays a role in the two theorems. However, when we consider variable populations and impose the frequently used \textit{consistency} axiom from the literature \citep{thomson1990consistency,thomson2011consistency}, \autoref{thm:consistency} shows that local unanimity together with consistency immediately characterizes TTC. Consistency requires that if an allocation has been determined for a given economy and subsequently a subset of agents is removed along with their assignments, and in the housing market model we further require that the removed agents' assignments coincide with their endowments, then the mechanism should preserve the original assignments for the remaining agents in the reduced economy. Consistency thus enables the iterative application of local unanimity in every reduced economy, thereby recovering the outcome of TTC.
	
	So far, our results have focused on strict preferences. Our final result addresses weak preferences. Although most of the literature assumes strict preferences, weak preferences frequently arise in applications (e.g., kidney exchange) and have been studied by important papers (e.g., \citealp{bogomolnaia2004random,bogomolnaia2005strategy}). 	
	A common practical approach to handling weak preferences is to apply exogenously fixed tie-breakers to transform weak preferences into strict ones. \cite{ehlers2014top} provides the first characterization of TTC with fixed tie-breakers in the housing market model with weak preferences. In a variable population setting, \citeauthor{ehlers2014top} employs the axioms of individual rationality, consistency, strategy-proofness, non-bossiness, and weak Pareto efficiency. In the same setting, \autoref{thm:fixed:tiebreaker} provides a new characterization using local unanimity, consistency, strategy-proofness, and non-bossiness. Under weak preferences, strategy-proofness and non-bossiness cannot be replaced by group strategy-proofness, as these properties are no longer equivalent.
	
	The remainder of the paper is organized as follows. After discussing the related literature in this section, \autoref{section:model} introduces the housing market model and the TTC mechanism. \autoref{section:unanimity} defines the local unanimity condition. \autoref{section:fixed:population} characterizes TTC in the fixed population setting and presents \autoref{thm:top-one} and \autoref{thm:top-two}. \autoref{section:variablepopulation} characterizes TTC in the variable population setting and presents \autoref{thm:consistency}. Finally, \autoref{section:fixed:tie-breaker} characterizes TTC with fixed tie-breakers under weak preferences and presents \autoref{thm:fixed:tiebreaker}.
	
	\paragraph{Related literature} Since \cite{ma1994strategy}'s seminal characterization of TTC, many other characterizations have been developed. The reader may refer to \cite{afacan2024housing} for a comprehensive survey. Here, we discuss some results related to ours.
    
    \cite{takamiya2001coalition} proves that TTC is the unique mechanism satisfying individual rationality, ontoness (unanimity), and group strategy-proofness; this result is implied by \cite{ma1994strategy} since ontoness and group strategy-proofness together imply Pareto efficiency.
	\cite{miyagawa2002strategy} proves that a mechanism that is individually rational, strategy-proof, anonymous, and non-bossy is either TTC or the no-trade mechanism (where agents retain their endowments).
	\cite{fujinaka2018endowments} prove that TTC is the only mechanism satisfying individual rationality, strategy-proofness, and endowment-swapping-proofness, which requires that no pair of agents gains by swapping their endowments before running the mechanism.\footnote{Several studies have explored the direction of weakening strategy-proofness; see, for example, \cite{altuntas2023characterizations,coreno2025characterizing}.} 
	Recently, \cite{ekici2023pair} strengthens \cite{ma1994strategy}'s result by relaxing Pareto efficiency to pair efficiency, a considerably weaker condition that only rules out welfare-improving exchanges between two agents.  \cite{klaus2025characterization} provide a characterization without using individual rationality, showing that TTC is the unique mechanism satisfying pair efficiency, improvement-respecting, and strategy-proofness.

	The above results are obtained on the domain of unrestricted strict preferences. When preferences are restricted, these characterizations may no longer hold. For instance, when agents have single-peaked preferences, \cite{bade2019matching} offers a mechanism distinct from TTC that satisfies \citeauthor{ma1994strategy}'s axioms. In contrast, when agents have single-dipped preferences, \cite{tamura2023object} proves that \citeauthor{ma1994strategy}'s result still holds and \cite{hu2024characterization} show that \citeauthor{ekici2023pair}'s result also remains valid.  

    Compared with the results discussed above, our approach employs a new axiom that is particularly natural in the housing market model. Notably, previous characterizations of TTC rely on at least three axioms, whereas our results require only two. For characterizing TTC with fixed tie-breakers under weak preferences, \autoref{thm:fixed:tiebreaker} uses four axioms, fewer than the five axioms employed by \citet{ehlers2014top}. In fact, we show that local unanimity, strategy-proofness, and non-bossiness together imply the individual rationality and weak Pareto efficiency conditions used by \citeauthor{ehlers2014top}.

   Local unanimity is a new axiom, but it has parallels in the literature. 
	An analogous concept in two-sided matching is the \textit{mutual best} axiom, which requires that if a pair of agents from distinct sides most prefer each other, they should be matched. \cite{toda2006monotonicity} shows that, in the marriage model, any set-valued solution satisfying individual rationality, mutual best, and Maskin monotonicity is a subset of the set of stable matchings.\footnote{\cite{klaus2011competition} examines mutual best in the roommate problem. \cite{morrill2013alternative} uses mutual best and other axioms to characterize TTC in the school choice model.} \cite{takagi2010impossibility} introduce \textit{respecting 2-unanimity}, which strengthens mutual best by additionally requiring that any agent preferring to remain unmatched should indeed be unmatched. They prove a negative result: no strategy-proof mechanism respects 2-unanimity. Generalizing this,  \cite{takamiya2013coalitional} studies a coalition formation model that subsumes the marriage model and defines \textit{coalitional unanimity}, which requires that if all members of a coalition rank the coalition at the top of their preferences, then the coalition should form.\footnote{\cite{rodriguez-alvarez2009strategyproof} introduce an axiom similar to  coalitional unanimity called \textit{top coalition}.} \citeauthor{takamiya2013coalitional} proves that if a strategy-proof mechanism exists that satisfies coalitional unanimity, then whenever a strictly core stable coalition structure exists, it must be unique and the mechanism must select it. In the marriage model, because the strictly core stable coalition structure is not unique, the result of \cite{takagi2010impossibility} follows. Coalitional unanimity is similar in spirit to local unanimity. However, our result is independent of \cite{takamiya2013coalitional}, because the housing market model is not a special case of the coalition formation model. In the housing market model, a group of agents who exchange endowments in a cycle may be viewed as a coalition. But because there are multiple ways for members within a coalition to exchange their endowments, specifying a coalition structure is insufficient to describe an allocation.

	Finally, in a very recent paper, \cite{kamada2025robust} study dropout problems in the housing market model and independently introduce an axiom called \textit{$k$-unanimity}. This axiom requires that, when the size of trading cycles is restricted to at most $ k $, every cycle that respects the size restriction and gives each involved agent his most preferred object must be implemented. When $k$ is at least as large as the number of agents, $k$-unanimity is equivalent to local unanimity. However,  \cite{kamada2025robust} focus on the case where the size restriction is nontrivial. They show that $k$-unanimity is incompatible with strategy-proofness and propose a mechanism that outperforms TTC when dropouts are possible. They do not provide any characterization results.

	\section{The housing market model} \label{section:model}
	
	We present the housing market model with a fixed population. 	
	Let $I=\{1,2,\ldots,n\}$ be a finite set of agents where $n\ge 3$. Let $O=\{o_1,o_2,\ldots,o_n\}$ be a finite set of indivisible objects such that, for each $ i\in I $, $o_i$ denotes $ i $'s endowment. Each $ i $ holds a strict preference relation $\succ_i$, which is a linear order of objects. We write $o\succsim_i o'$ to indicate that either $o\succ_i o'$ or $o=o'$. Let $top(\succ_i)$ denote the top choice in $\succ_i$. That is, $top(\succ_i)\succ_i o$ for every $o\in O\backslash \{top(\succ_i)\}$.
	%For any $ o\in O $, let $U(\succ_i,o)=\{o'\in O: o'\succ_i o\}$ be the upper contour set of $\succ_i $ at $o$.
	
	For any nonempty $J\subseteq I$, we define $\succ_{J}=(\succ_i)_{i\in J}$ to denote the preference profile of $J$. Therefore, $\succ_I$ is a preference profile of all agents. Fixing the set of agents with their endowments, a housing market economy is identified by a preference profile $\succ_{I}$.

	An allocation is a one-to-one mapping $\mu:I\rightarrow O$ where, for every $i\in I$, $\mu(i)$ denotes the object assigned to $i$. Let $\mathcal{M}_I$ denote the set of allocations.
	
	Given a preference profile $\succ_I$, an allocation $\mu \in \mathcal{M}_I$ is:
	\begin{itemize}
		\item \textbf{Individually rational} (IR) if, for every $i\in I$, $\mu(i)\succsim_i o_i$;
		
		\item \textbf{Pareto efficient} if there does not exist a different allocation $\mu'\in \mathcal{M}_I$ such that $\mu'(i)\succsim_i \mu(i)$ for all $i\in I$ and $\mu'(j)\succ_j \mu(j)$ for some $j\in I$;
		
		\item \textbf{Unanimously best} if, for any other allocation $\mu'\in \mathcal{M}_I$, $\mu(i)\succsim_i \mu'(i)$ for all $i\in I$.
	\end{itemize}
	
	Let $ \mathcal{P} $ denote the set of all strict preferences. Let $\mathcal{D}\subseteq \mathcal{P}$ denote a preference domain. Then, $ \D^n $ is the domain of preference profiles.
	
	Given a domain $ \D $, a mechanism is a mapping $\psi: \mathcal{D}^n \rightarrow \mathcal{M}_I$, which finds an allocation $\psi(\succ_I)$ for every $\succ_I \in \mathcal{D}^n$. For every $i\in I$, $\psi_i(\succ_I)$ denotes the object assigned to $i$ in $\psi(\succ_I)$.
	
	A mechanism $\psi$ is \textbf{individually rational}/\textbf{Pareto efficient} if, for every $\succ_I\in \mathcal{D}^n$, $\psi(\succ_I)$ is individually rational/Pareto efficient. A mechanism $\psi$ is \textbf{unanimous} if, for every $\succ_I\in \mathcal{D}^n$, whenever there is a unanimously best allocation $\mu$, $\psi(\succ_I)=\mu$.

	A mechanism $\psi$ is \textbf{strategy-proof} if, for all $\succ_I\in \mathcal{D}^n$ and all $i\in I$, there does not exist $\succ'_i\in \mathcal{D}\backslash \{\succ_i\}$ such that $\psi_i(\succ'_i,\succ_{I\backslash \{i\}}) \succ_i \psi_i(\succ_I)$.

	A mechanism $\psi$ is \textbf{group strategy-proof} if, for all $\succ_I\in \mathcal{D}^n$ and all $J\subseteq I$, there does not exist $\succ'_{J}\in \mathcal{D}^{|J|}$ such that, for all $i\in J$,  $\psi_i(\succ'_{J},\succ_{I\backslash J}) \succsim_i \psi_i(\succ_I)$, and for some $j\in J$, $\psi_j(\succ'_{J},\succ_{I\backslash J}) \succ_j \psi_j(\succ_I)$.
	
	The \textbf{Top Trading Cycles} (TTC) mechanism is defined as follows. For any preference profile, it finds an allocation in the following procedure:
	
	\begin{itemize}
		\item In the first step, let every agent point to the agent who owns his favorite object. In every generated cycle, let the agents exchange their endowments as indicated by the cycle. If an agent points to himself, he receives his own endowment. All agents involved in cycles are removed with their assignments.
		
		\item In every remaining step, repeat the above procedure for the remaining agents. Terminate when all agents are removed.
	\end{itemize}
	
	It is known that TTC is individually rational, Pareto efficient, and strategy-proof on any domain $ \D $. If $ \D=\mathcal{P} $, \cite{ma1994strategy} has shown that TTC is the unique mechanism satisfying the three properties. TTC is also group strategy-proof.

	\section{Local unanimity} \label{section:unanimity}
	
	Local unanimity is an axiom playing a central role in our characterizations of TTC. Recognizing that the housing market model represents a ``private good'' economy, we generalize the idea of unanimity to arbitrary subsets of agents. It says that if any subset of agents unanimously agrees that an exchange of their endowments gives them the best objects they can receive among all possible allocations in the original economy, then the exchange should be implemented.

	Formally, for every nonempty subset of agents $J$, let $O_{J}=\{o_i\}_{i\in J}$ denote the endowments of $J$. When $J$ allocates their endowments among themselves, it results in a \textbf{suballocation} of $O_{J}$, represented by a one-to-one mapping $\mu:J\rightarrow O_{J} $. Let $\mathcal{M}_{J}$ denote the set of suballocations of $O_{J}$.
	
	\begin{definition}\label{defn:local unanimous best}
		Given a preference profile $\succ_I$, for any nonempty $ J\subseteq I $, a suballocation $\mu\in \mathcal{M}_{J}$ is \textbf{unanimously best} for $J$ if, for all $i\in J$, $\mu(i)=top(\succ_i)$.
	\end{definition}

	\begin{definition}\label{defn:local unanimity}
		A mechanism $\psi$ is \textbf{locally unanimous} on a preference domain $ \D $, if, for every $\succ_I\in \mathcal{D}^n$ and every nonempty $J\subseteq I$, whenever there is a unanimously best suballocation $\mu\in \mathcal{M}_{J}$ for $ J $, $\psi_i(\succ_I)=\mu(i)$ for all $i\in J$.
	\end{definition}
	
	Clearly, local unanimity implies unanimity.
	There is a straightforward characterization of local unanimity: it is equivalent to requiring a mechanism ``coincide with the first step of TTC''. Specifically, for any nonempty $ J\subseteq I $, if  $\mu\in \mathcal{M}_{J}$ is unanimously best for $J$, then the members of $J$ must form one or several disjoint cycles in the first step of TTC such that, by clearing these cycles, we obtain the suballocation $ \mu $. Conversely, if a group $J$ form a cycle in the first step of TTC, by clearing the cycle, they must reach a unanimously best suballocation for the group. Thus, we obtain the following equivalence result.
	\begin{lemma}\label{lemma:first:step}
		A mechanism $\psi$ is locally unanimous on any $ \D $ if and only if, for every  $\succ_I\in \mathcal{D}^n$ and every $J\subseteq I$ who form a cycle in the first step of TTC, we have that, for all $i\in J$, $\psi_i(\succ_I)=top(\succ_i)$.
	\end{lemma}
	\begin{proof}
		(\underline{If}) For any  $\succ_I\in \mathcal{D}^n$, suppose that there exists a nonempty $ J\subseteq I $ such that some $\mu\in \mathcal{M}_{J}$ be unanimously best for $ J $. Thus, for all $ i\in J $, $ \mu(i)=top(\succ_i) $. Consider any $ i_1\in J $. If $ \mu(i_1)=o_{i_1} $, then $ i_1 $ must form a cycle with himself in the first step of TTC. By the condition in the lemma, $ \psi_{i_1}(\succ_I)= o_{i_1}$. If $ \mu(i_1)\neq o_{i_1} $, then there must exist $ i_2\in J $ such that $ \mu(i_1)=o_{i_2} $. Then, there must exist $ i_3\in J $ such that $ \mu(i_2)=o_{i_3} $, and so on. Since there are finite agents, there must exist a group of distinct agents, including $i_1$, who all belong to $ J $ such that they exchange their endowments in a cycle to reach their assignments in $\mu$. Such a cycle must appear in the first step of TTC. By the condition in the lemma, $ \psi_{i_1}(\succ_I)= \mu(i_1)$. Since $i_1$ is arbitrarily selected, we are equivalent to proving that, for all $i\in J$, $ \psi_i(\succ_I)= \mu(i)=top(\succ_i)$.
		
		(\underline{Only if}) For every  $\succ_I\in \mathcal{D}^n$  and every $J\subseteq I$ that form a cycle in the first step of TTC, the agents in $ J $ exchange their endowments and receive their favorite objects in the outcome of TTC. Thus, the resulting suballocation of their endowments must be unanimously best for them. Then, local unanimity requires that for all $i\in J$, $\psi_i(\succ_I)=top(\succ_i)$.
	\end{proof}
	
	Local unanimity has no restrictions on the groups of agents who do not have unanimously best suballocations of their objects. Thus, a locally unanimous mechanism does not need to be Pareto efficient or individually rational.

	\section{Characterization of TTC with a fixed population}\label{section:fixed:population}
	
	This section presents two characterizations of TTC in the fixed population setting. 	
	
	\subsection{Main results}
	
	We first consider the following condition on a preference domain $ \D $:
	
	\begin{itemize}
		\item[] \textbf{Top-one Richness}: For every $ o\in O $, there exists $\succ\in \mathcal{D}$ such that $top(\succ)=o$.
	\end{itemize}
	
	Top-one richness is a widely used condition in social choice; for instance, it is called \textit{minimal richness} in \cite{chatterji2013domains} and \cite{chatterji2019random}.  A domain satisfying Top-one Richness can be significantly smaller than the strict domain $ \mathcal{P} $. The domain $ \mathcal{P} $ contains $n!$ elements, whereas a domain satisfying Top-one Richness may contain as few as $n$ elements.
	
	We prove a lemma that is useful for the remaining results in the paper.

	\begin{lemma}\label{lemma:IR:PE}
		On any domain satisfying Top-one Richness, a locally unanimous and strategy-proof mechanism is individually rational.
	\end{lemma}
	
	\begin{proof}
		Suppose that a locally unanimous and strategy-proof mechanism $\psi$ finds an allocation that is not individually rational for some preference profile $\succ_I$. Then, there exists some $i\in I $ such that $o_i \succ_i \psi_i(\succ_I)$. Now, consider a preference relation $\succ'_i$ that ranks $o_i$ as top choice. Then, local unanimity requires that $\psi_i(\succ'_i,\succ_{-i})=o_i$. But this means that $i$ can manipulate $\psi$ in $\succ_I$ by reporting $\succ'_i$, which contradicts the strategy-proofness of $ \psi $.
	\end{proof}

We are ready to present the first theorem.

\begin{theorem}\label{thm:top-one}
	On any domain satisfying Top-one Richness, a mechanism is locally unanimous and group strategy-proof if and only if it is TTC.
\end{theorem}

Recall that a mechanism satisfying local unanimity must coincide with the first step of TTC in any preference profile. We prove that by imposing group strategy-proofness, the complete outcome of TTC is recovered. Our proof is constructive, explicitly explaining how the two axioms pin down the allocation for every preference profile.

\begin{proof}[Proof of \autoref{thm:top-one}]
	Clearly, TTC is locally unanimous and group strategy-proof. So, it is sufficient to prove the ``only if'' part. Let $\psi$ be a locally unanimous and group strategy-proof mechanism. Given any $\succ_I$, for any $k\ge 1$, let $I^k$ denote the set of agents involved in cycles in step $k$ of TTC.
	We want to prove that for every $\succ_I$, every step $k$ of TTC, and every $i\in I^k$, $\psi_i(\succ_I)=TTC_i(\succ_I)$. We prove this by induction on the steps of TTC.

	\textbf{Base case}. Consider any $\succ_I$. By \autoref{lemma:first:step}, local unanimity requires that for every $i\in I^1$, $\psi_i(\succ_I)=TTC_i(\succ_I)$.

	\textbf{Induction step}. Consider the same $\succ_I$. Suppose that for every $k\in \{1,2,\ldots,K-1\}$ and every $i\in I^k$, we have proved that $\psi_i(\succ_I)=TTC_i(\succ_I)$. We now consider step $K$.
	
	For every $i\in I^K$, $TTC_i(\succ_I)$ is $i$'s favorite object among $O_{I\backslash (I^1\cup I^2\cup \cdots \cup I^{K-1})}$ and $\psi_i(\succ_I)\in O_{I\backslash (I^1\cup I^2\cup \cdots \cup I^{K-1})}$. By \autoref{lemma:IR:PE}, $\psi$ is individually rational. So, for any $i\in I^K$ such that $TTC_i(\succ_I)=o_i$, it must be  that $\psi_i(\succ_I)=o_i=TTC_i(\succ_I)$. Let $I^*=\{i\in I^K: TTC_i(\succ_I) \succ_i o_i\}$. If $I^*$ is nonempty, the agents in $I^*$ must exchange endowments in cycles in step $K$ of TTC.  Consider any such cycle. If the agents in the cycle do not receive their TTC assignments, let them change to report preference relations that rank their TTC assignments as top choice. These preference relations are allowed by Top-one Richness. Then, local unanimity would require that all of the agents in the cycle must receive their TTC assignments. However, this means that $\psi$ is not group strategy-proof, a contradiction.
\end{proof}

The next subsection discusses independence of axioms in the theorem. When there are only three agents, \autoref{prop:three-agents} proves that group strategy-proofness in \autoref{thm:top-one} can be replaced by strategy-proofness.  However, when there are four or more agents, this relaxation is not possible. 
\autoref{example:single-peaked} in the next subsection constructs a mechanism different from TTC that is strategy-proof and locally unanimous on a domain satisfying Top-one Richness.

\begin{proposition} \label{prop:three-agents}
	On any domain satisfying Top-one Richness, when there are only three agents, a mechanism is locally unanimous and strategy-proof if and only if it is TTC.
\end{proposition}

\begin{proof}[\textbf{Proof of \autoref{prop:three-agents}} ]
	Because TTC is strategy-proof and locally unanimous, it is sufficient to prove the ``only if'' part. Let $\psi$ be a locally unanimous and strategy-proof mechanism on a domain satisfying Top-one Richness.  Consider any preference profile $\succ_I$. If there are two or more agents who are involved in cycles in the first step of TTC, then it is immediate that $\psi(\succ_I)=TTC(\succ_I)$. So, it is sufficient to consider the case that only one agent is involved in a cycle in the first step of TTC. Without loss of generality, let the agent be $ 1 $. Thus, $ top(\succ_1)=o_1 $. Because strategy-proofness and local unanimity imply individual rationality, $ \psi_1(\succ_I)=o_1=TTC_1(\succ_I) $.
	
	For the remaining two agents $ 2 $ and $ 3 $, if one of them prefers his own endowment over the other's endowment, individual rationality immediately requires that $\psi(\succ_I)=TTC(\succ_I)$. So, it is sufficient to consider the case that $ 2 $ and $ 3 $ prefer each other's endowment over their own. Consider a preference profile $(\succ_1,\succ'_2,\succ'_3)$ where $ 2 $ and $ 3 $ report each other's endowment as their top choice. Local unanimity requires that $\psi(\succ_1,\succ'_2,\succ'_3)=TTC(\succ_1,\succ'_2,\succ'_3)$. Next, consider the preference profile $(\succ_1,\succ_2,\succ'_3)$. Strategy-proofness requires that $\psi_2(\succ_1,\succ_2,\succ'_3)=\psi_2(\succ_1,\succ'_2,\succ'_3)=o_3$. Because $ 1 $ must receive $ o_1 $, we have $\psi_3(\succ_1,\succ_2,\succ'_3)=o_2$. Finally, strategy-proofness requires that $\psi_3(\succ_I)=\psi_3(\succ_1,\succ_2,\succ'_3)=o_2$. Given that 1 must receive $o_1$, $\psi_2(\succ_I)=o_3$. So, $\psi(\succ_I)=TTC(\succ_I)$.
\end{proof}

It is worth noting that \cite{ma1994strategy}'s characterization of TTC may not hold on a domain that satisfies only Top-one Richness. To see this, note that the domain of \textbf{single-peaked} preferences satisfies Top-one Richness.\footnote{The single-peaked domain specifies an order $\rhd$ of objects. A preference relation $\succ$ is \textbf{single-peaked with respect to $\rhd$} if there exists an object, denoted by $p(\succ)$, such that for each $o\in O\backslash \{p(\succ)\}$, $p(\succ)\succ o$, and for every distinct $o,o'\in O\backslash \{p(\succ)\}$, if $o\rhd o' \rhd p(\succ)$ or $ p(\succ) \rhd o' \rhd o$, then $o'\succ o$. The domain consists of all single-peaked preferences with respect to $ \rhd $.} On this domain,  \cite{bade2019matching} proposes a mechanism different from TTC that satisfies all axioms used by \citeauthor{ma1994strategy}. In fact, \citeauthor{bade2019matching}'s mechanism is group strategy-proof \citep{huang2024matching}. Thus, by \autoref{thm:top-one}, \citeauthor{bade2019matching}'s mechanism  is not locally unanimous (even when there are only three agents).

Our second theorem proves that group strategy-proofness in \autoref{thm:top-one} can be replaced by strategy-proofness if a stronger condition is imposed on the preference domain:
	\begin{itemize}
		\item[] \textbf{Top-two Richness}: For any two distinct objects $o,o'\in O$, there exists $\succ \in \D $ that ranks $o$ as the top choice and ranks $o'$ as the second choice.
	\end{itemize}

Top-two Richness has also been used in social choice; for instance, it is called the \textit{connectedness} condition in \cite{aswal2003dictatorial} and \cite{chatterji2011topsonly}. A domain satisfying Top-two Richness can contain as few as $n(n-1)$ elements.

	\begin{theorem}\label{thm:top-two}
		On any domain  satisfying Top-two Richness, a mechanism is locally unanimous and strategy-proof if and only if it is TTC.
	\end{theorem}

	\begin{proof}[Proof of \autoref{thm:top-two}]
		TTC is clearly locally unanimous and strategy-proof. So, it is sufficient to prove the ``only if'' part. Let $\psi$ be a locally unanimous and strategy-proof mechanism. Given any preference profile $\succ_I$, for any $k\ge 1$, let $I^k$ denote the set of agents involved in cycles in step $k$ of TTC.
		We want to prove that, for every $\succ_I$, every step $k$ of TTC, and every $i\in I^k$, $\psi_i(\succ_I)=TTC_i(\succ_I)$. We prove this by induction on the steps of TTC.

		\textbf{Base case.} By \autoref{lemma:first:step}, local unanimity implies that for every $\succ_I$ and every $i\in I^1$, $\psi_i(\succ_I)=TTC_i(\succ_I)$.
		
		\textbf{Induction step.} Suppose that for every $\succ_I$ where TTC has at least $K$ steps, we have proved that, for every $k\in \{1,2,\ldots,K-1\}$ and every $i\in I^k$,  $\psi_i(\succ_I)=TTC_i(\succ_I)$. That is, $\psi$ coincides with the first $K-1$ steps of TTC for every $\succ_I$ that admits at least $K$ steps of TTC. We then want to prove that, for every such $\succ_I$ and every $i\in I^K$, $\psi_i(\succ_I)=TTC_i(\succ_I)$.
		
		Consider any $\succ_I$ where TTC has at least $K$ steps. By the induction assumption, for every $i\in I^1\cup I^2\cup \cdots \cup I^{K-1}$, $\psi_i(\succ_I)=TTC_i(\succ_I)$. So, for every $i\in I^K$, $TTC_i(\succ_I)$ is $i$'s favorite object among $O_{I\backslash (I^1\cup I^2\cup \cdots \cup I^{K-1})}$ and $\psi_i(\succ_I)\in O_{I\backslash (I^1\cup I^2\cup \cdots \cup I^{K-1})}$. By \autoref{lemma:IR:PE}, $\psi$ is individually rational. So, for every $i\in I^K$ such that $TTC_i(\succ_I)=o_i$, it must be that $\psi_i(\succ_I)=o_i=TTC_i(\succ_I)$. Let $I^*=\{i\in I^K: TTC_i(\succ_I) \succ_i o_i\}$. If $I^*$ is nonempty, the agents in $I^*$ must exchange endowments in cycles in step $K$ of TTC.
		Without loss of generality, let $I^*=\{1,2,\ldots,m\}$. For every $i\in I^*$, let $\succ'_i$ denote a preference relation that ranks $TTC_i(\succ_I)$ as the top choice and $o_i$ as the second choice. By Top-two Richness, this preference relation must exist. Now, we inductively consider the following preference profiles and finally obtain that, for every $i\in I^*$, $\psi_i(\succ_I)=TTC_i(\succ_I)$.
		
		\textbf{Step 1}: We consider $\succ'_I=(\succ'_1,\succ'_2,\ldots,\succ'_m, \succ_{m+1},\ldots,\succ_n)$. Because each $i\in I^*$ ranks $TTC_i(\succ_I)$ as the top choice, they must be involved in cycles in step 1 of TTC in $\succ'_I$. Local unanimity of $\psi$ requires that, for each $i\in I^*$, $\psi_i(\succ'_I)=TTC_i(\succ_I)$.
		
		\textbf{Step 2}: For each $i\in I^*$, we consider $\succ'_{[i]}=(\succ'_{I^*\backslash i}, \succ_i, \succ_{I\backslash I^*})$. That is, $\succ'_{[i]}$ is obtained from $\succ'_I$ by changing $i$'s preferences from $\succ'_i$ to $\succ_i$. In $\succ'_{[i]}$, $I^1,I^2,\ldots,I^{K-1}$ must form the same cycles in the first $K-1$ steps of TTC as they do in $\succ_I$. By the induction assumption, for every $a\in I^1\cup I^2\cup \cdots \cup I^{K-1}$, $\psi_a(\succ'_{[i]})=TTC_a(\succ_I)$. Therefore, for each $i\in I^*$, $TTC_i(\succ_I) \succsim_i \psi_i(\succ'_{[i]})$. Since $\psi_i(\succ'_I)=TTC_i(\succ_I)$, strategy-proofness of $\psi$ requires that $ \psi_i(\succ'_{[i]})=TTC_i(\succ_I)$. Because $TTC_i(\succ_I)$ is the endowment of some $j\in I^*$ and $\succ'_j$ reports $TTC_j(\succ_I)$ as the top choice and $o_j$ as the second choice, individual rationality of $\psi$ requires that $ \psi_j(\succ'_{[i]})=TTC_j(\succ_I)$. Inductively, for every $j\in I^*$ that is involved in a cycle with $i$ in the procedure of TTC in $\succ'_{[i]}$, individual rationality of $\psi$ requires that $ \psi_j(\succ'_{[i]})=TTC_j(\succ_I)$. For the remaining agents in $I^*$, because they are involved in cycles in step 1 of TTC in $\succ'_{[i]}$, local unanimity of $\psi$ requires that, for each such agent $b$, $ \psi_b(\succ'_{[i]})=TTC_b(\succ_I)$.
		
		\textbf{Step 3}: For each two distinct $i,j\in I^*$, we consider $\succ'_{[i,j]}=(\succ'_{I^*\backslash i}, \succ_i, \succ_j, \succ_{I\backslash I^*})$. That is, $\succ'_{[i,j]}$ is obtained from $\succ'_{[i]}$ by changing $j$'s preferences from $\succ'_j$ to $\succ_j$ and is obtained from $\succ'_{[j]}$ by changing $i$'s preferences from $\succ'_i$ to $\succ_i$. As before, for every $a\in I^1\cup I^2\cup \cdots \cup I^{K-1}$, since they are involved in cycles in the first $K-1$ steps of TTC, $\psi_a(\succ'_{[i,j]})=TTC_a(\succ_I)$. So, for each $b \in I^*$,  $TTC_b(\succ_I) \succsim_b \psi_b(\succ'_{[i,j]})$. Then, since $ \psi_j(\succ'_{[i]})=TTC_j(\succ_I)$ and $ \psi_i(\succ'_{[j]})=TTC_i(\succ_I)$, strategy-proofness of $\psi$ requires that $ \psi_i(\succ'_{[i,j]})=TTC_i(\succ_I)$ and $ \psi_j(\succ'_{[i,j]})= TTC_j(\succ_I)$. For each $b \in I^*\backslash \{i,j\}$, if $b$ is involved in a cycle with $i$ or $j$ in the procedure of TTC in $\succ'_{[i,j]}$, individual rationality of $\psi$ requires that $ \psi_b(\succ'_{[i,j]})=TTC_b(\succ_I)$; otherwise, $b$ must be involved in a cycle in step 1 of TTC in $\succ'_{[i,j]}$, and local unanimity of $\psi$ requires that $ \psi_b(\succ'_{[i,j]})=TTC_b(\succ_I)$.
		
		\textbf{Step $\ell\ge 4$}: Inductively, for each $I'\subseteq I^*$ with $|I'|=\ell-1$, we consider $\succ'_{[I']}=(\succ'_{I^*\backslash I'}, \succ_{I'}, \succ_{I\backslash I^*})$. As before, for every $a\in I^1\cup I^2\cup \cdots \cup I^{K-1}$, $\psi_a(\succ'_{[I']})=TTC_a(\succ_I)$. So, for each $i\in I'$, $TTC_i(\succ_I) \succsim_i \psi_i(\succ'_{[I']})$. Because $\psi_i(\succ'_{[I'\backslash \{i\}]})=TTC_i(\succ_I)$, strategy-proofness of $\psi$ requires that, for each $i\in I'$, $ \psi_i(\succ'_{[I']})=TTC_i(\succ_I)$. For each remaining $b \in I^*\backslash I'$, if $b$ is involved in a cycle with some $i\in I'$ in the procedure of TTC in $\succ'_{[I']}$, individual rationality of $\psi$ requires that $ \psi_b(\succ'_{[I']})=TTC_b(\succ_I)$; otherwise, $b$ must be involved in a cycle in step 1 of TTC in $\succ'_{[I']}$, and local unanimity of $\psi$ requires that $ \psi_b(\succ'_{[I']})=TTC_b(\succ_I)$.
		
		When {$I'=I^*$}, $\succ'_{[I']}= \succ_I$. So, for every $i\in I^*$, $\psi_i(\succ_I)=TTC_i(\succ_I)$.
	\end{proof}

	Since the proof of \autoref{thm:top-two} is more involved than the proof of \autoref{thm:top-one}, we present an example to illustrate the induction step in the proof. It explains how the two axioms together pin down the allocation for a preference profile.
	
	\begin{example}\label{example:illustration}
		Suppose that $I=\{1,2,3\}$. Consider a preference profile $(\succ_1,\succ_2,\succ_3)$ where every agent most prefers $o_3$, and $1$ and $2$ prefer each other's endowment over their own. In the procedure of TTC, $3$ forms a self-cycle in step 1, and $\{1,2\}$ forms a cycle in step 2. Local unanimity only requires that $3$ receive his endowment. Below, we explain how strategy-proofness and local unanimity together require that $1$ and $2$ must exchange endowments. See illustrations in \autoref{table:proof:illustration}.
		
		Step 1: Consider $(\succ'_1,\succ'_2,\succ_3)$ in which $1$ and $2$ view each other's endowment as the top choice and their own endowment as the second choice. Local unanimity then requires that $1$ and $2$ must exchange endowments and $3$ receive $o_3$.
		
		Step 2: Consider $(\succ'_1,\succ_2,\succ_3)$ and $(\succ_1,\succ'_2,\succ_3)$. In each preference profile, local unanimity requires that $3$ receive $o_3$. In $(\succ_1,\succ'_2,\succ_3)$, by comparing it with $(\succ'_1,\succ'_2,\succ_3)$, strategy-proofness requires that $1$ must receive $o_2$. Then, individual rationality (implied by strategy-proofness and local unanimity) requires that $2$ must receive $o_1$. So, $1$ and $2$ must exchange endowments. The symmetric argument holds for $(\succ'_1,\succ_2,\succ_3)$.
		
		Step 3: Consider $(\succ_1,\succ_2,\succ_3)$. Local unanimity requires that $3$ must receive $o_3$. By comparing the preference profile with the two discussed in step 2, strategy-proofness requires that $1$ must receive $o_2$ and $2$ must receive $o_1$. So, the two agents must exchange endowments.

		\begin{table}[!hbt]
			\centering
			\begin{tabular}{ccc}
				$ \succ'_1 $ & $ \succ'_{2} $ & $\succ_3 $ \\ \hline
				$ o_2 $ &  $ o_1 $ & $ o_3 $ \\
				$ o_1$ & $o_2$ & $ o_1 $ \\
				$ o_3 $& $ o_3$ &  $ o_2$ \\
				
				\color{white} $ $ \\
				\multicolumn{3}{c}{$ \Downarrow $}
			\end{tabular}
			\quad $\Rightarrow$ \quad
			\begin{tabular}{ccc}
				$ \succ'_1 $ & $ \succ_{2} $ & $\succ_3 $ \\\hline
				$ o_2 $ &  $ o_3 $ & $ o_3 $ \\
				$ o_1$ & $o_1$ & $ o_1 $ \\
				$ o_3 $& $ o_2$ &  $ o_2$ \\
				\color{white} $ $ \\
				\multicolumn{3}{c}{$ \Downarrow $}
			\end{tabular}
			\\
			\begin{tabular}{ccc}
				\multicolumn{3}{c}{$ \color{white} \Downarrow $}\\
				$ \succ_1 $ & $ \succ'_{2} $ & $\succ_3 $ \\\hline
				$ o_3 $ &  $ o_1 $ & $ o_3 $ \\
				$ o_2$ & $o_2$ & $ o_1 $ \\
				$ o_1 $& $ o_3$ &  $ o_2$
			\end{tabular}
			\quad $\Rightarrow$ \quad
			\begin{tabular}{ccc}
				\multicolumn{3}{c}{$ \color{white} \Downarrow $}\\
				$\succ_1 $ & $ \succ_{2} $ & $\succ_3 $ \\\hline
				$ o_3 $ &  $ o_3 $ & $ o_3 $ \\
				$ o_2$ & $o_1$ & $ o_1 $ \\
				$ o_1 $& $ o_2$ &  $ o_2$
			\end{tabular}
			\caption{ \autoref{example:illustration}}\label{table:proof:illustration}
		\end{table}
	\end{example}
	
	It is not hard to verify that \cite{ma1994strategy}'s result holds on any domain satisfying Top-two Richness.\footnote{The readers may verify that the short proof presented by \cite{sethuraman2016alternative} for \citeauthor{ma1994strategy}'s result holds on any domain satisfying Top-two Richness.} \autoref{thm:top-two} is independent of \citeauthor{ma1994strategy}'s result because, in the absence of strategy-proofness, local unanimity does not imply individual rationality and Pareto efficiency, and vice versa.

	\subsection{Independence of axioms}\label{section:discussion}
	
	There exist mechanisms different from TTC that satisfy only one of the two axioms in the two theorems. 
		
	First, to see that local unanimity is necessary in both theorems, consider the no-trade mechanism that always assigns each agent his endowment. This mechanism is group strategy-proof but not locally unanimous on any domain satisfying Top-one Richness.

	Second, to see that strategy-proofness is necessary in  \autoref{thm:top-two}, consider the mechanism that lets the agents who form cycles in the first step of TTC receive their TTC assignments and lets the remaining agents receive their endowments. This mechanism is locally unanimous but not strategy-proof on any domain satisfying Top-one Richness.

	Last, to show that group strategy-proofness is necessary in \autoref{thm:top-one} when there are four or more agents, \autoref{example:single-peaked} constructs a locally unanimous and strategy-proof mechanism that violates group strategy-proofness on the single-peaked domain, which satisfies Top-one Richness. 
	
	In our model, group strategy-proofness is equivalent to the combination of strategy-proofness and non-bossiness. Thus, the  above constructions also show that strategy-proofness and non-bossiness each are indispensable for \autoref{thm:top-one}. The definition of non-bossiness is presented in \autoref{section:fixed:tie-breaker}.
	
	Moreover, the conditions on preference domains in the two theorems are also indispensable. Since the single-peaked domain in \autoref{example:single-peaked} violates Top-two Richness, the construction of a locally unanimous and strategy-proof mechanism different from TTC suggests that Top-two Richness is indispensable for  \autoref{thm:top-two}. 	
	
	\autoref{example:single-dipped} constructs a locally unanimous and group strategy-proof mechanism different from TTC on the \textbf{single-dipped} domain.\footnote{Given an order $\rhd$ of objects, a preference relation $\succ$ is \textbf{single-dipped with respect to $\rhd$} if there exists an object, denoted by $d(\succ)$, such that for each $o\in O\backslash \{d(\succ)\}$, $o\succ d(\succ)$, and for each distinct $o,o'\in O\backslash \{d(\succ)\}$, if $o\rhd o' \rhd d(\succ)$ or $ d(\succ) \rhd o' \rhd o$, then $o\succ o'$. The domain consists of all single-dipped preferences with respect to $\rhd$.} This domain violates Top-one Richness. Thus, it suggests that Top-one Richness is indispensable for  \autoref{thm:top-one}. 
	Interestingly, 
	\cite{tamura2023object} proves that \cite{ma1994strategy}'s result remains true on the single-dipped domain. Together with the fact that \citeauthor{ma1994strategy}'s result does not hold on the single-peaked domain, it suggests that \citeauthor{ma1994strategy}'s result and \autoref{thm:top-one} can hold on different domains.

	\begin{example}[Single-peaked preferences]\label{example:single-peaked}
		Let $I=\{1,2,\ldots,n\}$ with $n\ge 4$. Consider the single-peaked domain in which objects are ordered according to their indices: $o_1\rhd o_2 \rhd \cdots \rhd o_n$.
		
		Consider the following preference profiles $\succ_I=(\succ_1,\succ_2,\succ_3,\succ_4,\ldots,\succ_n)$:
		\[
		\begin{matrix}
			\succ_1 & \succ_2 & \succ_3 & \succ_4 & \cdots & \succ_n\\ \hline
			o_4 & \underline{o_1} & o_1 & \underline{o_4} &  & \underline{o_n} \\
			\underline{o_3}& o_2 & \underline{o_2} & \vdots &  & \vdots \\
			o_2 & o_3 & o_3 & \\
			o_1 & o_4 & o_4 &  \\
			\vdots & \vdots & \vdots
		\end{matrix}
		\]
		In words, 1,2,3 have the above preferences over $o_1,o_2,o_3,o_4$, and prefer these four objects over the others, while each remaining agent most prefers their own endowment.
		For each such $\succ_I$, TTC finds the allocation in which $1$ and $3$ exchange endowments, and the remaining agents receive their own endowments.
		
		However, we consider a mechanism $\psi$ such that: \begin{itemize}
			\item For each such $\succ_I$, $\psi$ finds the allocation in which the first three agents exchange endowments as in the cycle $1\rightarrow 3 \rightarrow 2 \rightarrow 1 $, and each remaining agent receives their own endowment. This allocation is underlined in the above table.
			
			\item For any other preference profile, $\psi$ finds the same allocation as TTC does.
		\end{itemize}

		It is clear that $\psi\neq TTC$, and $\psi$ is locally unanimous.
		Next, we show that $\psi$ is strategy-proof. We only need to verify each agent $i$'s incentive of manipulating preferences to make the preference profile switch between the above discussed $\succ_I$ and any other preference profile $(\succ'_i, \succ_{I\backslash \{i\}})$ that does not belong to the above discussed case.
		
		\begin{itemize}
			\item \textbf{Agent 1}: Consider $\succ_I$  and any $(\succ'_1,\succ_{I\backslash \{1\}})$ with $\succ'_1\neq \succ_1$.

			First, $1$ does not want to deviate in $\succ_I$, because given that the agents other than $1,2,3$ must receive their endowments,  $1$ has obtained the best object in $\succ_I$.
			
			Second, $1$ does not want to deviate in $(\succ'_1,\succ_{I\backslash \{1\}})$. If $\succ'_1$ ranks $o_1$ as top choice, then $1$ must obtain $o_1$ and does not want to deviate. If $1$ ranks $o_2$ as top choice, then $\psi$ finds the same allocation as the outcome of TTC in which
			$1$ and $2$  exchange endowments. So, $1$ does not want to deviate. If $1$ ranks $o_3$ as top choice, then $\psi$ finds the same allocation as the outcome of TTC in which
			$1$ and $3$  exchange endowments. So, $1$ does not want to deviate. If $1$ ranks any $o_m$ with $m\ge 4$ as top choice, since the agents other than $1,2,3$ must receive their endowments and $o_3\succ'_1 o_2 \succ'_1 o_1$, $\psi$ finds the same allocation as the outcome of TTC in which
			$1$ and $3$  exchange endowments. So, $1$ does not want to deviate.
			Thus, in all cases, $1$ does not want to deviate.
			
			\item \textbf{Agent 2}: Consider $\succ_I$ and any $(\succ'_2,\succ_{I\backslash \{2\}})$ with $\succ'_2\neq \succ_2$.
			
			First, $2$ does not want to deviate in $\succ_I$, because he has obtained the top choice $o_1$.
			
			Second, $2$ does not want to deviate in $(\succ'_2,\succ_{I\backslash \{2\}})$. If $\succ'_2$ ranks $o_2$ as top choice, he must obtain $o_2$ and does not want to deviate. If $\succ'_2$ ranks $o_3$ as top choice, then $\psi$ finds the same allocation as the outcome of TTC in which
			$1$ and $3$  exchange endowments and            $2$ obtains $o_2$. However,  since $o_2$ is between $o_1$ and $o_3$ in $\rhd$, $o_2 \succ'_2 o_1$. So, $2$ does not want to deviate to $\succ_2$. If $\succ'_2$ ranks any $o_m$ with $m\ge 4$ as top choice, since the agents other than $1,2,3$ must receive their endowments and $o_3\succ'_2 o_2 \succ'_2 o_1$, $\psi$ finds the same allocation as the outcome of TTC in which
			$1$ and $3$ exchange endowments and $2$ obtains $o_2$. Since $o_2 \succ'_2 o_1$, $2$ does not want to deviate to $\succ_2$.
			
			\item \textbf{Agent 3}: Consider $\succ_I$  and any $(\succ'_3,\succ_{I\backslash \{3\}})$ with $\succ'_3\neq \succ_3$.
			
			If $\succ'_3$ ranks $o_2$ as top choice, then for $(\succ'_3,\succ_{I\backslash \{3\}})$, $\psi$ finds the same allocation as the outcome of TTC in which $1,2,3$ exchange endowments as in the cycle $1\rightarrow 3 \rightarrow 2 \rightarrow 1 $. This allocation coincides with $\psi(\succ_I)$. So, $3$ does not want to deviate in either $\succ_I$ or $(\succ'_3,\succ_{I\backslash \{3\}})$ by reporting the other preference relation.
			
			If $\succ'_3$ ranks $o_3$ as top choice, then $3$ must receive $o_3$ in $(\succ'_3,\succ_{I\backslash \{3\}})$. So, $3$ does not want to deviate in either $\succ_I$ or $(\succ'_3,\succ_{I\backslash \{3\}})$ by reporting the other preference relation.
			
			If $\succ'_3$ ranks any $o_m$ with $m\ge 4$ as top choice, since the agents other than $1,2,3$ must receive their endowments and $o_3\succ'_3 o_2 \succ'_3 o_1$, $3$ must receive $o_3$ in $(\succ'_3,\succ_{I\backslash \{3\}})$. So, $3$ does not want to deviate in either $\succ_I$ or $(\succ'_3,\succ_{I\backslash \{3\}})$ by reporting the other preference relation.
			
			\item \textbf{Other agents}: For each $i\in I\backslash \{1,2,3\}$, since $i$ receives his top choice $o_i$ in $\succ_I$ and receives an object no worse than $o_i$ in any $(\succ'_i,\succ_{I\backslash \{i\}})$, $i$ does not want to deviate in either $\succ_I$ or $(\succ'_i,\succ_{I\backslash \{i\}})$ by reporting the other preference relation.
		\end{itemize}
		
	\end{example}

	\begin{example}[Single-dipped preference]\label{example:single-dipped}
		Without loss of generality, we consider the single-dipped domain in which all preferences are single-dipped with respect to the order $o_1\rhd o_2 \rhd \cdots \rhd o_n$. Then, for each preference relation $\succ$ within the domain, either $top(\succ)=o_1$ or $top(\succ)=o_n$.
		
		Consider a mechanism $\psi$ such that, if $1$ and $n$ rank each other's endowment as top choice, let the two agents exchange their endowments, and let the remaining agents keep their endowments; otherwise, let all agents keep their endowments. Obviously, $\psi\neq TTC$.
		
		$\psi$ is locally unanimous because in any preference profile, either $1$ and $n$ rank each other's endowment as top choice, or at least one of $1$ and $n$ ranks his own endowment as top choice. In both cases, $\psi$ satisfies those agents by letting them receive their top choice. Note that there never exists a unanimously best suballocation for any $J\subseteq I\backslash \{1,n\}$, because all agents must rank either $o_1$ or $o_n$ as top choice.

		$\psi$ is group strategy-proof because in every preference profile $\succ_I$: (1) If $top(\succ_1)=top(\succ_n)=o_1$, then only $1$ can change the allocation by misreporting preferences. But if $1$ changes to report $o_n$ as top choice, $1$ would receive $o_n$ and become worse off. So, $1$ does not want to misreport; (2) If $top(\succ_1)=top(\succ_n)=o_n$, then only $n$ can change the allocation by misreporting preferences. But if $n$ changes to report $o_1$ as top choice, $n$ would receive $o_1$ and become worse off. So, $n$ does not want to misreport; (3) If $top(\succ_1)=o_n$ and $top(\succ_n)=o_1$, then both of them can change the allocation by misreporting preferences. But if any of them changes to report his own endowment as top choice, he would receive his endowment and become worse off. So, they do not want to misreport.
	\end{example}

	\section{Characterization of TTC with variable populations}\label{section:variablepopulation}

	The previous section shows that when a domain satisfies Top-two Richness (Top-one Richness), (group) strategy-proofness empowers local unanimity to recover the outcome of TTC. These results may not hold when the domain violates the conditions. In this section, we present a characterization of TTC from a different perspective. In contrast to the previous section that considers a fixed population, this section considers variable populations and imposes a frequently used axiom in the literature called \textbf{consistency}. In essence, consistency requires that a mechanism consistently solve economies with different scales of populations. More precisely, it stipulates that if an allocation has been determined for a given economy and subsequently a subset of agents is removed alongside their respective assignments, then the mechanism should preserve the original assignments for the remaining agents in the reduced economy. However, when applying this concept to the housing market model, we only consider the scenarios where the group of removed agents is assigned the group's own endowments. This specification ensures that the reduced economy remains a valid instance of the housing market model. Leveraging the power of consistency, we can apply local unanimity to both a housing market economy and any of its reduced economies, thereby recovering the outcome of TTC.

	Formally, $I$ is now called the grand set of agents. Let $\mathcal{I}=2^I\backslash \{\emptyset\}$ denotes the set of all potential populations. Every $I'\in \mathcal{I}$ is called a \textbf{market} and every preference profile $\succ_{I'}$ is called an \textbf{economy} in the market $ I' $. Now, a preference domain is a profile $\mathcal{D}=\{\mathcal{D}_{I'}\}_{I'\in \mathcal{I}}$ where $\mathcal{D}_{I'}$ denotes the domain for market $I'$. Given a domain $ \D $, a mechanism $\psi$ specifies an allocation for every economy  $\succ_{I'}$ in every market $I'\in \mathcal{I}$, which is denoted by $\psi(\succ_{I'})$.  A mechanism is \textbf{locally unanimous} if it finds a locally unanimous allocation in every economy in every market.

	Given a market $I'\in \mathcal{I}$ and a preference relation $\succ\in \mathcal{D}_{I'} $, for any nonempty $J\subseteq I'$, let $\succ|_{J}$ denote the restriction of $\succ$ to $O_{J}$. That is, $\succ|_{J}$ is a linear order of $O_{J}$ that preserves the relative rankings of $O_{J}$ in $\succ$. A domain $\mathcal{D}$ is called \textbf{consistent} if, for every  $I'\in \mathcal{I}$, every $\succ\in \mathcal{D}_{I'} $, and every nonempty $J\subseteq I'$,  $\succ|_{J}\in \mathcal{D}_{J}$. The strict domain $ \mathcal{P} $ is consistent. The single-peaked domain and the single-dipped domain are also consistent if the preference relations in each market are single-peaked/single-dipped with respect to a fixed order of objects.
	
	Given an economy $\succ_{I'} $ in any market $ I' $, for any nonempty $J\subseteq I'$, $\{\succ_i|_{J}\}_{i\in J}$ is called a \textbf{reduced economy} of $\succ_{I'}$.
	A mechanism $ \psi $ is called \textbf{consistent} on a consistent domain $ \D $ if, for every $I'\in \mathcal{I}$, every $\succ_{I'}\in \mathcal{D}^{|I'|}_{I'}$, and every nonempty $J\subseteq I$ such that $\{\psi_i(\succ_{I'})\}_{i\in I'\backslash J}=O_{I'\backslash J}$,
	\begin{center}
		$\psi_i(\succ_{I'})=\psi_i(\{\succ_i|_{J}\}_{i\in J})$ for every $i\in J$.
	\end{center}

	\begin{theorem}\label{thm:consistency}
		On any consistent preference domain, a mechanism is locally unanimous and consistent if and only if it is TTC.
	\end{theorem}
	
	\begin{proof}[\normalfont \textbf{Proof of  \autoref{thm:consistency}}]
		It is easy to see that TTC is consistent on any consistent preference domain. TTC is also locally unanimous. So, we only need to prove the ``only if'' part.
		Let $\psi$ be a mechanism that satisfies the two axioms.		
		Consider any economy $\succ_{I'} $ in any market $I'\in \mathcal{I}$. Let $I^k$ denote the set of agents involved in cycles in step $k$ of TTC. Local unanimity requires that, for every $i\in I^1$, $\psi_i(\succ_{I'})=TTC_i(\succ_{I'})$. Clearly, $\{\psi_i(\succ_{I'})\}_{i\in I^1}=O_{I^1}$.
		We then consider the reduced economy $\{\succ_i|_{I'\backslash I^1}\}_{i\in I'\backslash I^1}$. For every $i\in I^2$, $\succ_i|_{I'\backslash I^1}$ must rank $TTC_i(\succ_{I'})$ as top choice. Then, local unanimity requires that, for every $i\in I^2$, $\psi_i(\{\succ_i|_{I'\backslash I^1}\}_{i\in I'\backslash I^1})=TTC_i(\{\succ_i|_{I'\backslash I^1}\}_{i\in I'\backslash I^1})$. Because $\psi$ and TTC are consistent, for every $i\in I^2$,  $\psi_i(\succ_{I'})=\psi_i(\{\succ_i|_{I'\backslash I^1}\}_{i\in I'\backslash I^1})$ and $TTC_i(\succ_{I'})=TTC_i(\{\succ_i|_{I'\backslash I^1}\}_{i\in I'\backslash I^1})$. So, for every $i\in I^2$, $ \psi_i(\succ_{I'})=TTC_i(\succ_{I'}) $. Then, we can consider the reduced economy $\{\succ_i|_{I'\backslash (I^1\cup I^2)}\}_{i\in I'\backslash (I^1\cup I^2)}$ and repeat the above argument to conclude that, for every $i\in I^3$, $\psi_i(\succ_{I'})=TTC_i(\succ_{I'})$. Inductively, we can prove that, for every $i\in I^k$, $\psi_i(\succ_{I'})=TTC_i(\succ_{I'})$. So, $\psi=TTC$.
	\end{proof}
	\textbf{Independence of axioms}
	
	There exist mechanisms different from TTC that satisfy only one of the two axioms.

	\begin{itemize}
		\item (Local unanimity)
		The no-trade mechanism that always assigns each agent his endowment is consistent on any consistent domain, but it is not locally unanimous.

		\item (Consistency) The mechanism that lets the agents who form cycles in the first step of TTC receive their TTC assignments but lets the remaining agents receive their endowments is locally unanimous but not consistent on a consistent domain.
	\end{itemize}
	
	\section{Characterization of TTC with fixed tie-breakers under weak preferences}\label{section:fixed:tie-breaker}
	
	So far, we have restricted attention to strict preferences. However, in applications such as kidney exchange, we cannot rule out the possibility that agents may be indifferent between some objects. To deal with weak preferences, a frequently used method in applications is to apply exogenously fixed tie-breakers to transform weak preferences into strict preferences. \cite{ehlers2014top} provides the first characterization of TTC with fixed tie-breakers in the housing market model with weak preferences. This section shows that local unanimity can be used to provide an alternative characterization.

	Following \cite{ehlers2014top}, we consider a housing market model with variable populations, which has been introduced in \autoref{section:variablepopulation}. Different from previous sections that discuss varied conditions on the preference domain, we focus on the universal domain that allows agents to have any preferences. But as \cite{ehlers2014top}, we assume that each $ i $ is never indifferent between his endowment and any other object.
	
	Formally, in the grand market $ I $, for each $ i\in I $, let $ \mathcal{U}_i $ denote his preference domain, consisting of all complete and transitive preference relations over $ O $ that do not treat $ o_i $ as indifferent to any other object. Then, $ \mathcal{U}=\times_{i\in I} \mathcal{U}_i $ denotes the universal domain of preference profiles. In each market $ J\subsetneq I $, the preference domain for each $i\in J$ consists of the preferences in $\mathcal{U}_i$ restricted to $O_J$. For each $ \succsim_i\in \mathcal{U}_i $ and each distinct $ o,o'\in O$, we write $ o\succ_i o' $ if $ o $ is strictly better than $ o' $, and we write $ o\sim_i o' $ if they are indifferent for $ i $.

	TTC with fixed tie-breakers is defined as follows. For each $ i $, a strict order $ \rhd_i $ of the objects in $ O $ is specified. In each market $ I'\in \mathcal{I} $, for each $ i\in I' $ and each preference relation $ \succsim_i $ that $ i $ may hold, $ \succ^{\rhd_i}_i $ represents a strict transformation of $ \succsim_i $ such that, for each distinct $ o,o'\in O_{I'}$, if $ o\succ_i o' $, or $ o\sim_i o $ and $ o\rhd_i o' $, then $ o \succ^{\rhd_i}_i o $. Given a profile of tie-breakers $ \rhd=(\rhd_i)_{i\in I} $, $ TTC^\rhd $ is a mechanism such that, in each market $ I'\in \mathcal{I}$, for each preference profile $ \succsim_{I'} $, $ TTC^\rhd(\succsim_{I'})=TTC(\succsim^\rhd_{I'}) $, where  $ \succsim^\rhd_{I'}= (\succ_i^{\rhd_i})_{i\in I'}$.
	
	To accommodate weak preferences, in any market $I'\in \mathcal{I}$, for any preference profile $\succsim_{I'}$ and any nonempty $ J\subseteq I' $, we say that a suballocation $\mu\in \mathcal{M}_{J}$ is \textbf{unanimously best} for $J$ if, for every $i\in J$, $\mu(i)$ is the unique best object among $O_{I'}$. A mechanism $\psi$ is \textbf{locally unanimous} if in every market $I'$, for every preference profile, whenever there exists a unanimously best suballocation for any $J\subseteq I'$, each $i\in J$ receives their unique best object. This definition reduces to \autoref{defn:local unanimity} under strict preferences.

	A mechanism $\psi$ satisfies \textbf{non-bossiness} if in every market $I'\in \mathcal{I}$, for every preference profile $\succsim_{I'}$, every $i\in I'$, and every $\succsim'_i$ in $i$'s preference domain, if $\psi_i(\succsim_{I'})=\psi_i(\succsim'_i,\succsim_{I'\backslash i})$, then $\psi(\succsim_{I'})=\psi(\succsim'_i,\succsim_{I'\backslash i})$. That is, no agent can change the outcome of the mechanism without changing his own assignment.
	
	It is well known that under strict preferences, group strategy-proofness is equivalent to the combination of strategy-proofness and non-bossiness. However, this equivalence breaks down under weak preferences. Our characterization result utilizes non-bossiness, in addition to strategy-proofness.

	\begin{theorem}\label{thm:fixed:tiebreaker}
		On the universal domain under weak preferences, (1) for any profile of tie-breakers $\rhd$, $TTC^\rhd$ satisfies local unanimity, consistency, strategy-proofness, and non-bossiness; (2) if a mechanism $\psi$ satisfies local unanimity, consistency, strategy-proofness, and non-bossiness, then there exists a profile of tie-breakers $\rhd$ such that $\psi$ is welfare equivalent to $TTC^\rhd$.\footnote{Two mechanisms are welfare equivalent if in every economy, every agent views his assignments in two mechanisms as indifferent.}
	\end{theorem}
	
	The proof for the first part of  \autoref{thm:fixed:tiebreaker} is straightforward. The crux is to prove the second part. Our proof is built on the result of \cite{ehlers2014top}.  \cite{ehlers2014top} characterizes TTC with fixed tie-breakers by individual rationality, consistency, strategy-proofness, non-bossiness, and weak Pareto efficiency. An allocation in an economy is \textbf{weak Pareto efficient} if it is impossible to make every agent strictly better off. Since the universal domain $\mathcal{U}_i$ for each $i$ satisfies Top-one Richness, local unanimity and strategy-proofness imply individual rationality. Below, \autoref{lemma:weakPE} proves that local unanimity, strategy-proofness, and non-bossiness together imply weak Pareto efficiency. Note that these three axioms do not imply Pareto efficiency, because strategy-proofness and non-bossiness together no longer imply group strategy-proofness.

	\begin{lemma}\label{lemma:weakPE}
		On the universal domain under weak preferences, local unanimity, strategy-proofness, and non-bossiness together imply weak Pareto efficiency.
	\end{lemma}

	\begin{proof}
		Suppose that a locally unanimous, strategy-proof, and non-bossy mechanism $\psi$ finds an allocation that is not weak Pareto efficient for some preference profile $\succsim_{I'}$ in some market $I'\in \mathcal{I}$. Then, there exists $\mu \in \mathcal{M}_{I'}$ such that $\mu(i) \succ_i \psi_i(\succsim_{I'})$ for all $i \in I'$. For each $i\in I'$, let $\succsim_i'$ be a preference relation that ranks $\mu(i)$ as the unique best object, ranks $\psi_i(\succsim_{I'})$ above the other objects that are indifferent with $\psi_i(\succsim_{I'})$ in $\succsim_i$, and preserves the relative ordering of the other objects in $\succsim_i$. Without loss of generality, assume that $I'=\{1,2,\ldots,m\}$.
		
		Consider the preference profile $(\succsim'_1,\succsim_{I'\backslash \{1\}})$. Since $\{o\in O_{I'}: o\succ'_1 \psi_1(\succsim_{I'}) \}=\{o\in O_{I'}: o\succ_1 \psi_1(\succsim_{I'}) \} $ and $\{o\in O_{I'}: o\succsim'_1 \psi_1(\succsim_{I'}) \}=\{o\in O_{I'}: o\succ_1 \psi_1(\succsim_{I'}) \}\cup \{\psi_1(\succsim_{I'}\} $, strategy-proofness of $\psi$ requires that $\psi_1(\succsim'_1,\succsim_{I'\backslash \{1\}})=\psi_1(\succsim_{I'})$.\footnote{If $\psi_1(\succsim'_1,\succsim_{I'\backslash \{1\}})\succ'_1 \psi_1(\succsim_{I'})$, then 1 can manipulate $\psi$ at $\succsim_{I'}$ by reporting $\succsim'_1$. Conversely, if $\psi_1(\succsim_{I'})\succ'_1 \psi_1(\succsim'_1,\succsim_{I'\backslash \{1\}})$, then 1 can manipulate $\psi$ at $(\succsim'_1,\succsim_{I'\backslash \{1\}})$ by reporting $\succsim_1$.} Then, by non-bossiness, $\psi(\succsim'_1,\succsim_{I'\backslash \{1\}})=\psi(\succsim_{I'})$. Repeating the same argument for agents $2,3,\ldots,m$, we obtain $\psi(\succsim'_{I'})=\psi(\succsim_{I'})$. However, since every $i\in I'$ ranks $\mu(i)$ as the unique best object in $\succsim'_{I'}$, by local unanimity, $\psi(\succsim'_{I'})=\mu$. This is a contradiction.
	\end{proof}
	
	\begin{proof}[\normalfont \textbf{Proof of  \autoref{thm:fixed:tiebreaker}}]
		\underline{(1)} For any profile of tie-breakers $\rhd$, \cite{ehlers2014top} has proved that $TTC^\rhd$ satisfies consistency, strategy-proofness, and non-bossiness. So, we only need to prove that $TTC^\rhd$ satisfies local unanimity. In any market $I'\in \mathcal{I}$, for any preference profile $\succsim_{I'}$, if there exists a nonempty $ J\subseteq I' $ such that a suballocation $\mu\in \mathcal{M}_{J}$ is unanimously best for $J$, then in the strict preference profile $\succsim^\rhd_{I'}$, $\mu$ must be unanimously best for $J$. Since $TTC^\rhd$ satisfies local unanimity on the strict domain, for each $i\in J$, $TTC^\rhd(\succsim_{I'})=\mu(i)$. Thus, $TTC^\rhd$ satisfies local unanimity under weak preferences.
		
		\underline{(2)} Since the domain $\mathcal{U}_i$ for each $i$ satisfies Top-one Richness, by \autoref{lemma:IR:PE}, local unanimity and strategy-proofness imply individual rationality. \autoref{lemma:weakPE} has proved that local unanimity, strategy-proofness, and non-bossiness imply weak Pareto efficiency. Then, by the result of \cite{ehlers2014top}, $\psi$ is welfare equivalent to $TTC^\rhd$ for some $\rhd$.
	\end{proof}
	
	\textbf{Independence of axioms}
	
	For each axiom of \autoref{thm:fixed:tiebreaker}, there exists a mechanism that violates the axiom but satisfies the remaining axioms:

	\begin{itemize}
		\item (Local unanimity) The no-trade mechanism that always assigns each agent his endowment satisfies all axioms but local unanimity.

		\item (Consistency) Select two TTC with fixed tie-breakers $TTC^\rhd$ and $TTC^{\rhd'}$ such that, for some economy $\succsim_I$, $TTC^\rhd(\succsim_I)\neq TTC^{\rhd'}(\succsim_I)$. Define a mechanism $\psi$ that coincides with $TTC^\rhd$ in the market $I$ and coincides with $TTC^{\rhd'}$ in the remaining markets. Then, $\psi$ violates consistency but satisfies the remaining axioms.

		\item (Strategy-proofness) Suppose that $|I|=3$. For agent 1, define two different tie-breakers $\rhd_1:3,2,1$ and $\rhd'_1:2,3,1$. For the other two agents $i=2,3$, define the same tie-breaker $\rhd_i:3,2,1$. Let $\rhd=(\rhd_1,\rhd_2,\rhd_3)$ and $\rhd'=(\rhd'_1,\rhd_2,\rhd_3)$. Define a mechanism $\psi$ such that: (1) $\psi$ coincides with $TTC^\rhd$ in all markets $I'\subsetneq I$; (2) in the grand market $I$,  $\psi$ coincides with $TTC^\rhd$ in all economies expect that, for preference profiles $ (\succsim_1,\succsim_2,\succsim_3) $ where $\succsim_1$ ranks $o_2$ and $o_3$ both as top choices, and $\succsim_2$ and $\succsim_3$ both rank $o_1$ as the only top choice,    $\psi(\succsim_1,\succsim_2,\succsim_3)=TTC^{\rhd'}(\succsim_1,\succsim_2,\succsim_3)$. For such a preference profile $(\succsim_1,\succsim_2,\succsim_3)$, under $TTC^{\rhd'}$, 1 and 2 exchange endowments, while under $TTC^\rhd$, 1 and 3 exchange endowments.
		
		To see that $\psi$ is not strategy-proof, consider the preference profile $(\succsim_1,\succsim'_2,\succsim_3)$ in which $\succsim'_2$ ranks $o_1$ and $o_3$ both as top choice. Then, $\psi_2(\succsim_1,\succsim'_2,\succsim_3)=TTC^{\rhd}_2(\succsim_1,\succsim'_2,\succsim_3)=o_2$. However, $\psi_2(\succsim_1,\succsim_2,\succsim_3)=TTC^{\rhd'}_2(\succsim_1,\succsim_2,\succsim_3)=o_1$. So, 2 can manipulate $\psi$ at $(\succsim_1,\succsim'_2,\succsim_3)$ by reporting $\succsim_2$.
		
		$\psi$ is locally unanimous because in every economy its found allocation coincides with the outcome of some TTC with fixed tie-breakers. 
		
		$\psi$ is consistent because whenever $\psi$ deviates from $TTC^\rhd$ in the market $I$, $1$ and $2$ must exchange endowments, and thus after removing either $\{1,2\}$ or $\{3\}$, the resulting allocation in the reduced economy must coincide with the outcome of $TTC^\rhd$. 
		
		$\psi$ is non-bossy because, in any market $I'\subsetneq I$, $\psi$ is obviously non-bossy, while in the market $I$, no agent can change the allocation without changing his own assignment.
		
		\item (Non-bossiness) Suppose that $|I|=3$. For the three agents, define the tie-breakers $\rhd_1:3,2,1$, $\rhd_2:1,3,2$, and $\rhd_3:1,2,3$. Let $\rhd=(\rhd_1,\rhd_2,\rhd_3)$. Define a mechanism $\psi$ such that: (1) $\psi$ coincides with $TTC^\rhd$ in all markets $I'\subsetneq I$; (2) in the grand market $I$,  $\psi$ coincides with $TTC^\rhd$ in all economies expect that, for preference profiles $ (\succsim_1,\succsim_2,\succsim_3) $ where $\succsim_1$ ranks both $o_2$ and $o_3$ as top choices,  $\succsim_2$ ranks $o_3$ above $o_2$, and $\succsim_3$ ranks $o_1$ as a top choice,  $\psi$ assigns $o_2$ to 1, assigns $o_3$ to 2, and assigns $o_1$ to 3.
		
		To see that $\psi$ is bossy, consider the preference profile $(\succsim'_1,\succsim_2,\succsim_3)$ where $\succsim'_1$ ranks $o_2$ as the only top choice, $\succsim_2$ ranks $o_1$ as the only top choice and ranks $o_3$ above $o_2$, and $\succsim_3$ ranks $o_1$ as the only top choice. Under $\psi$, 1 and 2 will exchange endowments. But if $1$ changes to report $\succsim_1$ that ranks both $o_2$ and $o_3$ as top choices, the three agents will exchange endowments as in the cycle $1\rightarrow 2 \rightarrow 3 \rightarrow 1$. So, $1$ can change the allocation without changing his own assignment.
		
		$\psi$ is locally unanimous because whenever $\psi$ deviates from $TTC^\rhd$ in the market $I$, there does not exist a unanimously best suballocation for any subset of agents, so that local unanimity has no restriction. 
		
		$\psi$ is consistent because whenever $\psi$ deviates from $TTC^\rhd$ in the market $I$, the three agents exchange endowments, so that the reduced economy is null.

		$\psi$ is strategy-proof because, when $\psi(\succsim_I) \neq TTC^\rhd(\succsim_I)$ and $\succsim_2 $ ranks $ o_3 $ as a top choice, all three agents receive their top choices in $ \psi(\succsim_I) $ and thus have no incentive to deviate to any other preference profile. When $\psi(\succsim_I) \neq TTC^\rhd(\succsim_I)$ and $\succsim_2$ ranks $ o_1 $ as the only top choice, agents 1 and 3 still obtain their top choices, while agent 2 cannot improve by deviating from $\succsim_2$. In cases where $\psi$ coincides with $TTC^\rhd$, no agent can profitably deviate to a preference profile such that $\psi$ still coincides with $TTC^\rhd$, and it  can be verified that no agent has an incentive to deviate from any other preference profile to a profile $\succsim_I$ such that $\psi(\succsim_I) \neq TTC^\rhd(\succsim_I)$.

	\end{itemize}
	
	\setlength{\bibsep}{0pt plus 0.3ex}
	\bibliography{reference}

@article{morrill2013alternative,
	title={An alternative characterization of top trading cycles},
	author={Morrill, Thayer},
	journal={Economic Theory},
	volume={54},
	number={1},
	pages={181--197},
	year={2013},
	publisher={Springer}
}

@article{chatterji2013domains,
	title = {On Domains That Admit Well-Behaved Strategy-Proof Social Choice Functions},
	author = {Chatterji, Shurojit and Sanver, Remzi and Sen, Arunava},
	year = 2013,
	month = may,
	journal = {Journal of Economic Theory},
	volume = {148},
	number = {3},
	pages = {1050--1073},
	publisher = {Academic Press},
	issn = {0022-0531},
	doi = {10.1016/j.jet.2012.10.005},
	urldate = {2025-10-21},
	lccn = {3}
}

@article{chatterji2019random,
	title = {Random Mechanism Design on Multidimensional Domains},
	author = {Chatterji, Shurojit and Zeng, Huaxia},
	year = 2019,
	month = jul,
	journal = {Journal of Economic Theory},
	volume = {182},
	pages = {25--105},
	publisher = {Academic Press},
	issn = {0022-0531},
	doi = {10.1016/j.jet.2019.04.003},
	urldate = {2025-10-21},
	lccn = {3}
}

@article{chatterji2011topsonly,
	title = {Tops-Only Domains},
	author = {Chatterji, Shurojit and Sen, Arunava},
	year = 2011,
	month = feb,
	journal = {Economic Theory},
	volume = {46},
	number = {2},
	pages = {255--282},
	publisher = {Springer},
	issn = {0938-2259},
	doi = {10.1007/s00199-009-0509-2},
	urldate = {2025-10-21},
	lccn = {3}
}

@article{aswal2003dictatorial,
	title={Dictatorial domains},
	author={Aswal, Navin and Chatterji, Shurojit and Sen, Arunava},
	journal={Economic Theory},
	volume={22},
	number={1},
	pages={45--62},
	year={2003},
	publisher={Springer}
}

@article{huang2024matching,
	title={Matching within Limited Distance},
	author={Huang, Zhen and Tian, Guoqiang},
	year={2024},
	journal={working paper}
}

@article{bogomolnaia2004random,
	title={Random matching under dichotomous preferences},
	author={Bogomolnaia, Anna and Moulin, Herv{\'e}},
	journal={Econometrica},
	volume={72},
	number={1},
	pages={257--279},
	year={2004},
	publisher={Wiley Online Library}
}

@article{bogomolnaia2005strategy,
	title={Strategy-proof assignment on the full preference domain},
	author={Bogomolnaia, Anna and Deb, Rajat and Ehlers, Lars},
	journal={Journal of Economic Theory},
	volume={123},
	number={2},
	pages={161--186},
	year={2005},
	publisher={Elsevier}
}

@book{arrow1951social,
	title     = {Social Choice and Individual Values},
	author    = {Arrow, Kenneth J.},
	year      = {1951},
	publisher = {Yale University Press}
}

@article{feng2024characterizing,
	title={Characterizing the typewise top-trading-cycles mechanism for multiple-type housing markets},
	author={Feng, Di and Klaus, Bettina and Klijn, Flip},
	journal={Games and Economic Behavior},
	volume={146},
	pages={234--254},
	year={2024},
	publisher={Elsevier}
}

@article{ehlers2014top,
	title={Top trading with fixed tie-breaking in markets with indivisible goods},
	author={Ehlers, Lars},
	journal={Journal of Economic Theory},
	volume={151},
	pages={64--87},
	year={2014},
	publisher={Elsevier}
}

@article{thomson2011consistency,
	title={Consistency and its converse: an introduction},
	author={Thomson, William},
	journal={Review of Economic Design},
	volume={15},
	number={4},
	pages={257--291},
	year={2011},
	publisher={Springer}
}

@incollection{thomson1990consistency,
	title={The consistency principle},
	author={Thomson, William},
	booktitle={Game theory and applications},
	pages={187--215},
	year={1990},
	publisher={Elsevier}
}

@article{shapley1974cores,
	title={On cores and indivisibility},
	author={Shapley, Lloyd and Scarf, Herbert},
	journal={Journal of Mathematical Economics},
	volume={1},
	number={1},
	pages={23--37},
	year={1974},
	publisher={Elsevier}
}

@article{bade2019matching,
	title={Matching with single-peaked preferences},
	author={Bade, Sophie},
	journal={Journal of Economic Theory},
	volume={180},
	pages={81--99},
	year={2019},
	publisher={Elsevier}
}

@article{tamura2023object,
	title={Object reallocation problems with single-dipped preferences},
	author={Tamura, Yuki},
	journal={Games and Economic Behavior},
	volume={140},
	pages={181--196},
	year={2023},
	publisher={Elsevier}
}

@article{sethuraman2016alternative,
	title={An alternative proof of a characterization of the TTC mechanism},
	author={Sethuraman, Jay},
	journal={Operations Research Letters},
	volume={44},
	number={1},
	pages={107--108},
	year={2016},
	publisher={Elsevier}
}

@article{ekici2023pair,
	author = {Ekici, {\"O}zg{\"u}n},
	title = {Pair-efficient reallocation of indivisible objects},
	journal = {Theoretical Economics},
	volume = {19},
	number = {2},
	year = {2024},
	pages = {551-564},
	url = {https://econtheory.org/ojs/index.php/te/article/view/20240551/0}
}

@article{ma1994strategy,
	title={Strategy-proofness and the strict core in a market with indivisibilities},
	author={Ma, Jinpeng},
	journal={International Journal of Game Theory},
	volume={23},
	pages={75--83},
	year={1994},
	publisher={Springer}
}

@article{miyagawa2002strategy,
	title={Strategy-proofness and the core in house allocation problems},
	author={Miyagawa, Eiichi},
	journal={Games and Economic Behavior},
	volume={38},
	number={2},
	pages={347--361},
	year={2002},
	publisher={Elsevier}
}

@article{takamiya2001coalition,
	title={Coalition strategy-proofness and monotonicity in Shapley--Scarf housing markets},
	author={Takamiya, Koji},
	journal={Mathematical Social Sciences},
	volume={41},
	number={2},
	pages={201--213},
	year={2001},
	publisher={Elsevier}
}

@article{fujinaka2018endowments,
	title={Endowments-swapping-proof house allocation},
	author={Fujinaka, Yuji and Wakayama, Takuma},
	journal={Games and Economic Behavior},
	volume={111},
	pages={187--202},
	year={2018},
	publisher={Elsevier}
}

@article{gibbard1973manipulation,
  title={Manipulation of Voting Schemes: A General Result},
  author={Gibbard, Allan},
  journal={Econometrica},
  volume={41},
  number={4},
  pages={587--601},
  year={1973}
}

@article{satterthwaite1975strategy,
	title={Strategy-proofness and Arrow's conditions: Existence and correspondence theorems for voting procedures and social welfare functions},
	author={Satterthwaite, Mark Allen},
	journal={Journal of Economic Theory},
	volume={10},
	number={2},
	pages={187--217},
	year={1975},
	publisher={Elsevier}
}

@article{altuntas2023characterizations,
	title = {Some Characterizations of {{Generalized Top Trading Cycles}}},
	author = {Altunta{\c s}, A{\c c}elya and Phan, William and Tamura, Yuki},
	year = {2023},
	month = sep,
	journal = {Games and Economic Behavior},
	volume = {141},
	pages = {156--181},
	issn = {08998256},
	doi = {10.1016/j.geb.2023.05.004},
	urldate = {2024-09-10},
	lccn = {3}
}

@article{hu2024characterization,
	title = {Characterization of {{Top Trading Cycles}} with Single-Dipped Preferences},
	author = {Hu, Xinquan and Zhang, Jun},
	year = {2024},
	month = aug,
	journal = {Economics Letters},
	volume = {241},
	pages = {111822},
	issn = {0165-1765},
	doi = {10.1016/j.econlet.2024.111822},
	urldate = {2024-06-26}
}

@article{klaus2025characterization,
	title = {A Characterization of the Top-Trading-Cycles Mechanism for Housing Markets via Respecting-Improvement},
	author = {Klaus, Bettina and Klijn, Flip and Sethuraman, Jay},
	year = {2025},
	month = feb,
	journal = {Economics Letters},
	volume = {247},
	pages = {112145},
	issn = {0165-1765},
	doi = {10.1016/j.econlet.2024.112145},
	urldate = {2025-03-13},
	lccn = {4}
}

@article{coreno2025characterizing,
	title = {Characterizing {{TTC}} via Endowments-Swapping-Proofness and Truncation-Proofness},
	author = {Coreno, Jacob and Feng, Di},
	year = {2025},
	month = feb,
	journal = {Economics Letters},
	volume = {247},
	pages = {112159},
	issn = {0165-1765},
	doi = {10.1016/j.econlet.2024.112159},
	urldate = {2025-03-13},
	lccn = {4}
}

@article{toda2006monotonicity,
	title = {Monotonicity and {{Consistency}} in {{Matching Markets}}},
	author = {Toda, Manabu},
	year = {2006},
	month = apr,
	journal = {International Journal of Game Theory},
	volume = {34},
	number = {1},
	pages = {13--31},
	issn = {0020-7276},
	doi = {10.1007/s00182-005-0002-5},
	urldate = {2025-07-01},
	lccn = {4}
}

@article{takamiya2013coalitional,
	title = {Coalitional Unanimity versus Strategy-Proofness in Coalition Formation Problems},
	author = {Takamiya, Koji},
	year = {2013},
	month = feb,
	journal = {International Journal of Game Theory},
	volume = {42},
	number = {1},
	pages = {115--130},
	issn = {0020-7276},
	doi = {10.1007/s00182-012-0318-x},
	urldate = {2025-06-19},
	lccn = {4}
}

@article{takagi2010impossibility,
	title = {An Impossibility Theorem for Matching Problems},
	author = {Takagi, Shohei and Serizawa, Shigehiro},
	year = {2010},
	month = jul,
	journal = {Social Choice and Welfare},
	volume = {35},
	number = {2},
	pages = {245--266},
	issn = {0176-1714},
	doi = {10.1007/s00355-009-0439-8},
	urldate = {2025-07-01},
	lccn = {4}
}

@article{rodriguez-alvarez2009strategyproof,
	title = {Strategy-Proof Coalition Formation},
	author = {{Rodr{\'i}guez-{\'A}lvarez}, Carmelo},
	year = {2009},
	month = nov,
	journal = {International Journal of Game Theory},
	volume = {38},
	number = {3},
	pages = {431--452},
	issn = {0020-7276},
	doi = {10.1007/s00182-009-0162-9},
	urldate = {2025-09-29},
	lccn = {4}
}

@article{kamada2025robust,
	title={Robust Exchange},
	author={Kamada, Yuichiro and Yasuda, Yosuke},
	journal={Available at SSRN 5500459},
	year={2025}
}

@article{klaus2011competition,
	title = {Competition and Resource Sensitivity in Marriage and Roommate Markets},
	author = {Klaus, Bettina},
	year = {2011},
	month = may,
	journal = {Games and Economic Behavior},
	volume = {72},
	number = {1},
	pages = {172--186},
	issn = {0899-8256},
	doi = {10.1016/j.geb.2010.07.004},
	urldate = {2025-09-29},
	lccn = {3}
}

@article{afacan2024housing,
  title = {Housing Markets since {{Shapley}} and {{Scarf}}},
  author = {Afacan, Mustafa O{\u g}uz and Hu, Gaoji and Li, Jiangtao},
  year = 2024,
  month = apr,
  journal = {Journal of Mathematical Economics},
  volume = {111},
  pages = {102967},
  issn = {0304-4068},
  doi = {10.1016/j.jmateco.2024.102967},
  urldate = {2024-03-11}
}

\end{document}